\UseRawInputEncoding
\documentclass[11pt, a4paper]{article}
\usepackage[ margin=1in]{geometry}
\usepackage[affil-it]{authblk}
\usepackage{setspace}
\usepackage[sort&compress,numbers]{natbib}
\usepackage[english]{babel}
\usepackage{graphicx}
\usepackage{mathptmx}
\usepackage{verbatim}
\usepackage{graphicx}
\usepackage{amsmath}
\usepackage{amssymb}
\usepackage{amsfonts}
\usepackage{mathrsfs}
\usepackage{amsmath} 
\usepackage{romannum}
\usepackage{color}
\usepackage{graphicx}
\usepackage{bm} 
\usepackage{amssymb}
\usepackage{xspace}
\usepackage{color}
\usepackage{float}
\usepackage{bbold}
\usepackage{tabu}
\usepackage{textcomp}
\usepackage{dcolumn}
\usepackage{graphicx}
\usepackage{dcolumn}
\usepackage{bm}
\usepackage{epstopdf}
\usepackage{xr}
\usepackage{amsmath}
\usepackage{color}
\usepackage{xcolor}
\usepackage[normalem]{ulem}
\usepackage{wasysym}
\usepackage{siunitx}
\usepackage{physics}
\usepackage{lineno}
\usepackage{gensymb}
\usepackage[T1]{fontenc}
\usepackage{booktabs}
\usepackage[section]{placeins}
\usepackage{textcomp}
\usepackage{caption}
\usepackage[colorlinks=true, linkcolor=black, urlcolor=blue, citecolor=blue]{hyperref}
\usepackage{blindtext}
\usepackage[absolute]{textpos}
\usepackage{makecell}

\makeatletter

\newcommand{\flin}{$f_{\mathrm{res}}^{\mathrm{lin}}$}

\makeatother

\begin{document}
\captionsetup[figure]{labelfont={bf},labelsep=period,name={Fig.}}

\pagenumbering{arabic}
\title{\vspace{-2cm}\normalfont \textbf{Kerr non-linearity enhances the response of a graphene Josephson bolometer}}

\renewcommand\Authfont{\small}

\author[1*]{Joydip Sarkar}
\author[1]{Krishnendu Maji}
\author[1,2]{Abhishek Sunamudi}
\author[1]{Heena Agarwal}
\author[1]{Priyanka Samanta}
\author[1]{Anirban Bhattacharjee}
\author[1]{Rishiraj Rajkhowa}
\author[1]{Meghan P Patankar}

\affil[1]{Department of Condensed Matter Physics and Materials Science, Tata Institute of Fundamental Research, Homi Bhabha Road, Mumbai 400005, India.}
\affil[2]{Indian Institute of Science Education and Research Kolkata, West Bengal 741246, India.}

\author[3]{Kenji Watanabe}
\affil[3]{Research Center for Functional Materials,
National Institute for Materials Science, 1-1 Namiki, Tsukuba 305-0044, Japan.}

\author[4]{Takashi Taniguchi}
\affil[4]{International Center for Materials Nanoarchitectonics,
National Institute for Materials Science,  1-1 Namiki, Tsukuba 305-0044, Japan.}

\author[1*]{Mandar M. Deshmukh}
\affil[*]{\textnormal{joydips581@gmail.com, deshmukh@tifr.res.in}}

\date{}
\maketitle

\begin{abstract}
Highly sensitive, broadband bolometers are of great interest because of their versatile usage for dark matter search, radio astronomy, material science, and qubit readouts in cQED experiments. There have been different realizations of bolometers using superconducting thin films, nanowires, quantum dots, and various $2$D materials in the recent past. The challenge is to have a single device that combines high sensitivity, broad bandwidth, a fast readout mechanism, and low noise. Here we demonstrate the first usage of a Josephson parametric amplifier (JPA) as a highly sensitive bolometer. Our key finding is the Kerr non-linearity of the JPA boosts the device's sensitivity. When the bolometer is biased in the non-linear regime, it enhances the sideband signals ($\sim100$ times), resulting in an order of magnitude improvement in sensitivity compared to the linear regime. In the non-linear biasing of the device, we achieve a NEP $\sim500$ aW/$\mathrm{\sqrt{Hz}}$. Our bolometer offers a fast detection scheme with a thermal time constant of $4.26$ $\mu$s and an intrinsic JPA time constant of $70$ ns. Our device's broadband and fast operation are key and new compared to previously studied graphene-based bolometers. In our device, the gate voltage tunability and the possibility of multiplexing combined with the sensitive bolometric performance offer an opportunity for integrated quantum sensor arrays. Our work demonstrates a way forward to enhance the performance of quantum sensors based on $2$D materials by leveraging the inherent non-linear response.
\end{abstract}
\clearpage 

Sensing single photons in the radio frequency (RF) range is challenging because of the low photon energies compared to THz or near-infrared (NIR). However, sensitive bolometers in radio frequency range are at the heart of radio astronomy and dark matter search experiments~\cite{paolucci_fully_2021,braine_extended_2020}. There are various types of bolometers utilizing different physical platforms, such as transition edge sensor (TES)~\cite{de_visser_fluctuations_2014}, microwave kinetic inductance detector (MKID)~\cite{pirro_advances_2017}, quantum dot-based~\cite{ikushima_photon-counting_2006}, superconducting qubit-based~\cite{inomata_single_2016,balembois_cyclically_2024}, graphene-based~\cite{yan_dual-gated_2012,efetov_fast_2018}, graphene Josephson junction-based~\cite{lee_graphene-based_2020,kokkoniemi_bolometer_2020,walsh_josephson_2021}, and superconducting nanowire-based (SNSPD)~\cite{oripov_superconducting_2023} sensors among others. The key performance metrics of any bolometer include noise equivalent power (NEP), response time ($\tau$), instantaneous frequency bandwidth, and operational frequency range. However, achieving optimal performance across all these metrics in a single device is challenging. Currently, the best noise equivalent power (NEP) for microwave photon is $10^{-22}$ W/$\sqrt{\text{Hz}}$ using a superconducting qubit-based bolometer~\cite{balembois_cyclically_2024}. However, this device operates within a narrow bandwidth of approximately $50$ MHz and requires a long integration time. Whereas, the fastest graphene-based bolometer operates with a very short thermal relaxation time of $35$ ps~\cite{efetov_fast_2018}. Nonetheless, its NEP is relatively modest at $10^{-12}$ W/$\sqrt{\text{Hz}}$. Therefore, efforts are required to improve all these aspects at a single-device level.

Over the past decade, single-photon bolometers based on graphene have gained popularity because of their high performance across a broad electromagnetic spectrum and their rapid response times. Graphene has many attractive electrical and thermal properties, rendering it a promising material for bolometry and calorimetry. At the charge neutrality, graphene has a vanishing density of states, resulting in small heat capacity and electron-to-phonon thermal coupling, which are highly desirable properties for bolometers~\cite{yan_dual-gated_2012}. It has been shown that graphene efficiently absorbs photons across a wide frequency spectrum starting from GHz to near infrared~\cite{efetov_fast_2018,cai_sensitive_2014,el_fatimy_epitaxial_2016}. The short electron-electron scattering time in graphene facilitates rapid equilibration of energy from absorbed photons~\cite{tielrooij_photoexcitation_2013,brida_ultrafast_2013}. Recently, superconductor-normal-superconductor (SNS) Josephson junctions have been used for bolometry harnessing the sensitivity of JJs~\cite{walsh_graphene-based_2017, lee_graphene-based_2020, kokkoniemi_bolometer_2020,walsh_josephson_2021,katti_hot_2023,huang_graphene_2024}. While these experiments have demonstrated highly sensitive operations, some utilize a low-frequency switching scheme to operate the device, which limits the bolometer's response time. Additionally, all these experiments show very narrow bandwidth operation and none of them have exploited the non-linear response of a JJ for sensing. Hence, there is scope for improving the sensitivity, response time, and bandwidth of these devices and we focus on these aspects in our work. 

To devise a highly sensitive, fast, and broadband bolometer, we employ a graphene-based JPA architecture. JPAs are widely used for quantum sensing and readout purposes in cQED experiments~\cite {aumentado_superconducting_2020}. Because of the extreme sensitivity of the JPA non-linearity and quantum noise-limited operation, such devices have the potential to sense at individual photon levels with minimal backaction~\cite{hatridge_dispersive_2011,sarkar_quantum-noise-limited_2022,butseraen_gate-tunable_2022}. While the previous realizations of graphene-based bolometers have exploited the temperature-dependent resistance, inductance, or switching events in JJs, the temperature-dependent non-linear Josephson inductance has been unexplored. In this work, we demonstrate a graphene JPA as a highly sensitive bolometer, in the first such application of JPA. We implement a separate heater line inside the device to inject RF signals onto the graphene flake. We measure the heater-modulated sideband response of the bolometer and extract the NEP and response time of the device experimentally.

\begin{figure*}
    \centering
    \includegraphics[width=15.5cm]{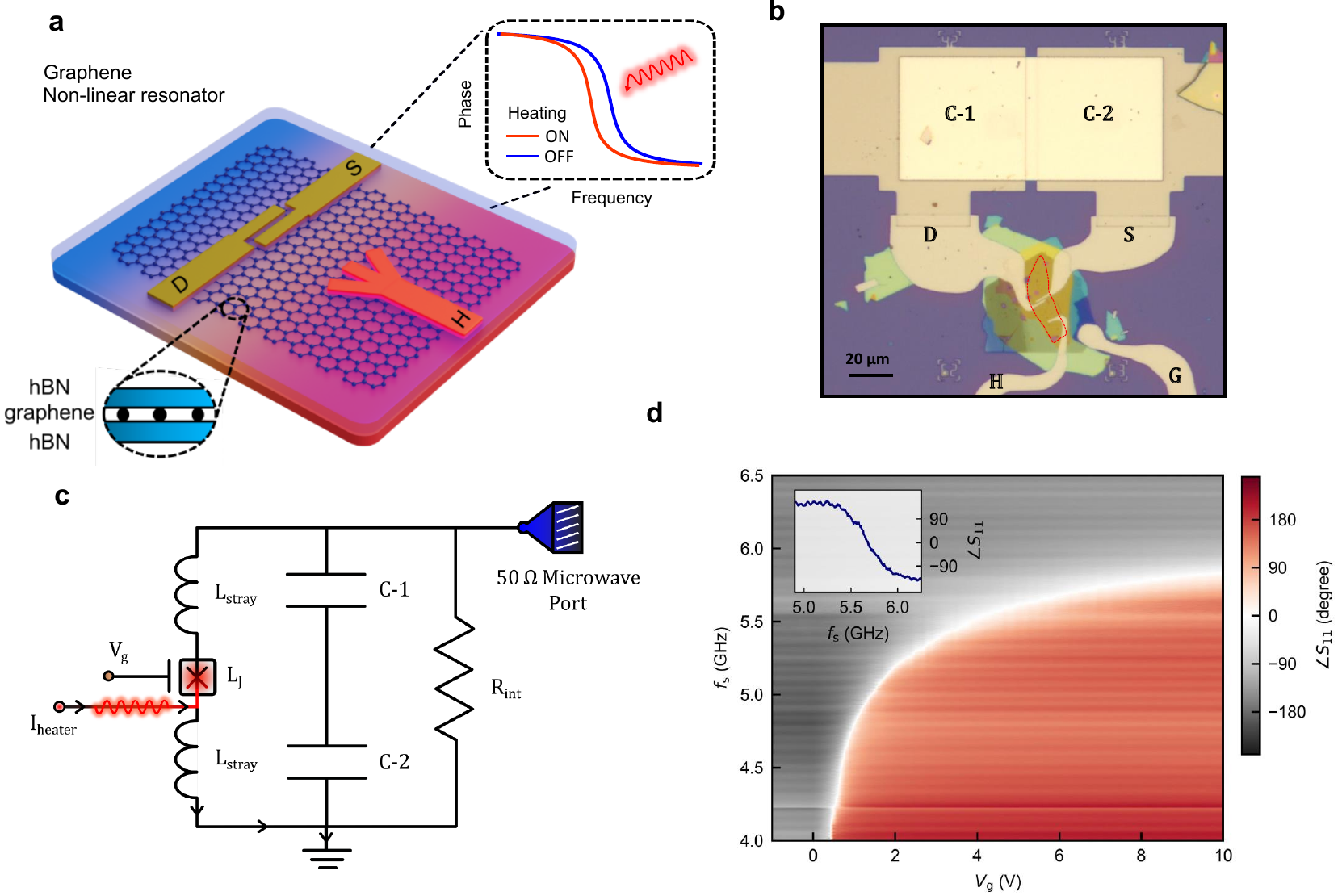}
    \caption{ \label{fig:bolodevimage} { \textbf{Device image, circuit model, and microwave response of the device.} 
        \textbf{a,}~Shows the schematic of our bolometer device, the JJ is made on an hBN-gr-hBN stack. We implement a separate heater line inside the device to inject the heating signals. In the image, S, D, G, and H imply the source, drain, gate, and heater electrodes, respectively.
		\textbf{b,}~Shows a zoomed optical micrograph of our device. The graphene JJ is in parallel to a series combination of two parallel plate capacitors C-$1$ and C-$2$ having identical dimensions of $60$x$60$ $\mu$m$^2$.  The red dashed line indicates the position of the sandwiched graphene flake. The heater line connects to the edge of the extended graphene flake which is substantially away from the junction. When the heater current flows it causes Joule heating in the extended graphene flake shared with the JJ; in turn, the generated heat propagates and increases the local temperature of the junction. The heater current drains to the ground plane of the CPW. 
            \textbf{c,}~Shows the equivalent lumped element circuit model of the JPA where $L_J$ is the junction inductance, $L_\mathrm{stray}$ is the stray inductance, C-$1$ and C-$2$ are the two parallel plate capacitors in series, $R_\mathrm{int}$ is to account for internal losses in the device and $V_g$ is the applied gate voltage to the JJ. The heater current is sent from a heater port and drains to the ground. The device is connected to a microwave port through a $50$ $\Omega$ matched environment for reflection-based measurements. 
                \textbf{d,}~Shows reflected phase ($\angle S_{11}$) of the bolometer device plotted as a function of signal frequency ($f_{\mathrm{s}}$) and applied gate voltage ($V_{\mathrm{g}}$). The linear resonance frequency of the device (\flin) gets tuned as a function of gating. The gate voltage tunes the linear resonance in a large frequency band of $4-5.7$ GHz. The $2\pi$ phase change at resonance indicates that the device is over coupled $(Q_{\mathrm{int}}\gg Q_{\mathrm{ext}})$ all along the electron doping side. The inset shows a line slice of the phase plot at $V_{\mathrm{g}}=6$ V.
    }}

\end{figure*}

Fig.~\ref{fig:bolodevimage}a-c shows the schematic, optical image, and circuit model of the bolometer device. The device contains a superconducting coplanar waveguide (CPW) where the central line is terminated to the ground through a lumped element LC resonator, made from a parallel plate capacitor and a graphene JJ which serves as an inductor.  The device is designed to have a low-quality factor ($\sim10$) using a directly coupled architecture that helps broadband and fast readout from the device. The JJ is made on an hBN-gr-hBN stack. We make a separate electrode close to the JJ ($10$ $\mu$m away) to inject heating signals onto the graphene flake. The graphene JJ and heater are on the same sheet of graphene that also serves as the bus for carrying excitations. 

The essential idea here is that for a JPA, the reflected signal's phase is a very sensitive parameter due to its nonlinear nature. Consequently, any minute heating effect causes a resonance frequency or phase shift in the device. The current-phase relation (CPR) of the graphene JJ, including the first non-linear term, can be approximated as
\begin{equation}\label{eqn:1}
    I_{\mathrm{s}}(\phi) \propto \frac{\sin{\phi}}{\sqrt{1-t_r\sin^2{(\phi/2)}}} \cong (\alpha\phi +\beta\phi^3).
\end{equation}
In Eq.~\eqref{eqn:1}, $I_{\mathrm{s}}$ is the supercurrent, $\phi$ is the phase difference across the junction, $\alpha$ and $\beta$ are the expansion coefficients, and $t_r$ is the averaged transparency factor for $N$ conducting channels in graphene. The Josephson inductance is given by $L_{\mathrm{J}}(\phi) = \frac{\hslash}{2e} \left(\frac{\partial I_{\mathrm{s}}}{\partial\phi}\right)^{-1}$. Hence, the non-linear inductance can be approximated as $L_{\mathrm{J}}(\phi)\cong L_0+L_2\phi^2$, where $L_0 $ and $L_2$ are the nonlinear inductor coefficients. This cubic nonlinearity in the CPR, known as the Kerr term, gives rise to a nonlinear junction inductance that is essential for parametric amplification. In our JPA bolometer device, we explore the effect of heating and the consequent phase shift in the device. When the heater signal is modulated, the resulting phase shift generates sidebands, which we use as markers for sensing.

\begin{figure*}

    \centering
	\includegraphics[width=15cm]{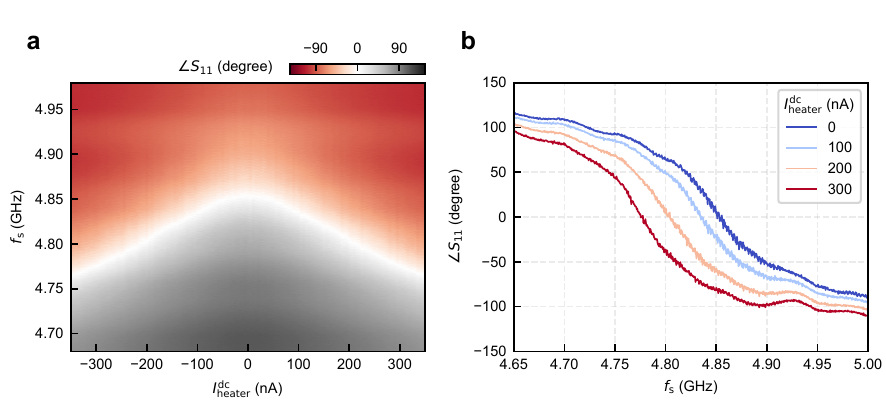}
    \caption{ \label{fig:boloDCphseshft}{\textbf{Resonance frequency shift of the bolometer due to heating.} 
    \textbf{a,}~Shows the low-power reflected phase ($\angle S_{11}$) of the device as a function of dc heater current ($I_{\mathrm{heater}}^{\mathrm{dc}}$) and microwave signal frequency ($f_{\mathrm{s}}$) at $V_{\mathrm{g}}=1$ V. The white color in the plot indicates the decreasing resonant feature of the device. The Joule heating caused by the heater current reduces the switching current ($I_c$) and increases the Josephson inductance of the junction. The decrease in the linear resonance frequency with heating is consistent with the \flin$\propto\sqrt{I_{\mathrm{c}}}$ prediction.
    \textbf{b,}~Shows the line slice of the color plot in Fig.~\ref{fig:boloDCphseshft}a for different values of the DC heater currents ($I_{\mathrm{heater}}^{\mathrm{dc}}$). 
}}
\end{figure*}

We begin with a basic microwave characterization of the device and then move to the bolometer measurements. We probe the device with low-power microwave signals and measure the reflected signals using a vector network analyzer (VNA). For DC characterization of the device see supplementary information section I. The measurements were done in a dilution fridge at $20$ mK temperature (see supplementary information section II for details on the setup). We check the gate tunability of the JPA and observe broad tunable resonance over a band of $\sim2$ GHz, see Fig.~\ref{fig:bolodevimage}d. Next, we proceed to do bolometer characterization, where we measure the reflected phase ($\angle S_{11}$) of the device as a function of signal frequency ($f_{\mathrm{s}}$) and applied dc heater current ($I_{\mathrm{heater}}^{\mathrm{dc}}$) at a fixed gate voltage ($V_{\mathrm{g}}=1$ V), see Fig.~\ref{fig:boloDCphseshft}a. The Joule heating caused by the heater current reduces the switching current ($I_c$) and increases the Josephson inductance of the junction. We observe a decrease in the linear resonance frequency with heating, consistent with the \flin$\propto\sqrt{I_{\mathrm{c}}}$ prediction. In Fig.~\ref{fig:boloDCphseshft}b we show the line slice of Fig.~\ref{fig:boloDCphseshft}a at different heater currents.

\begin{figure*}
    \centering
	\includegraphics[width=15.5cm]{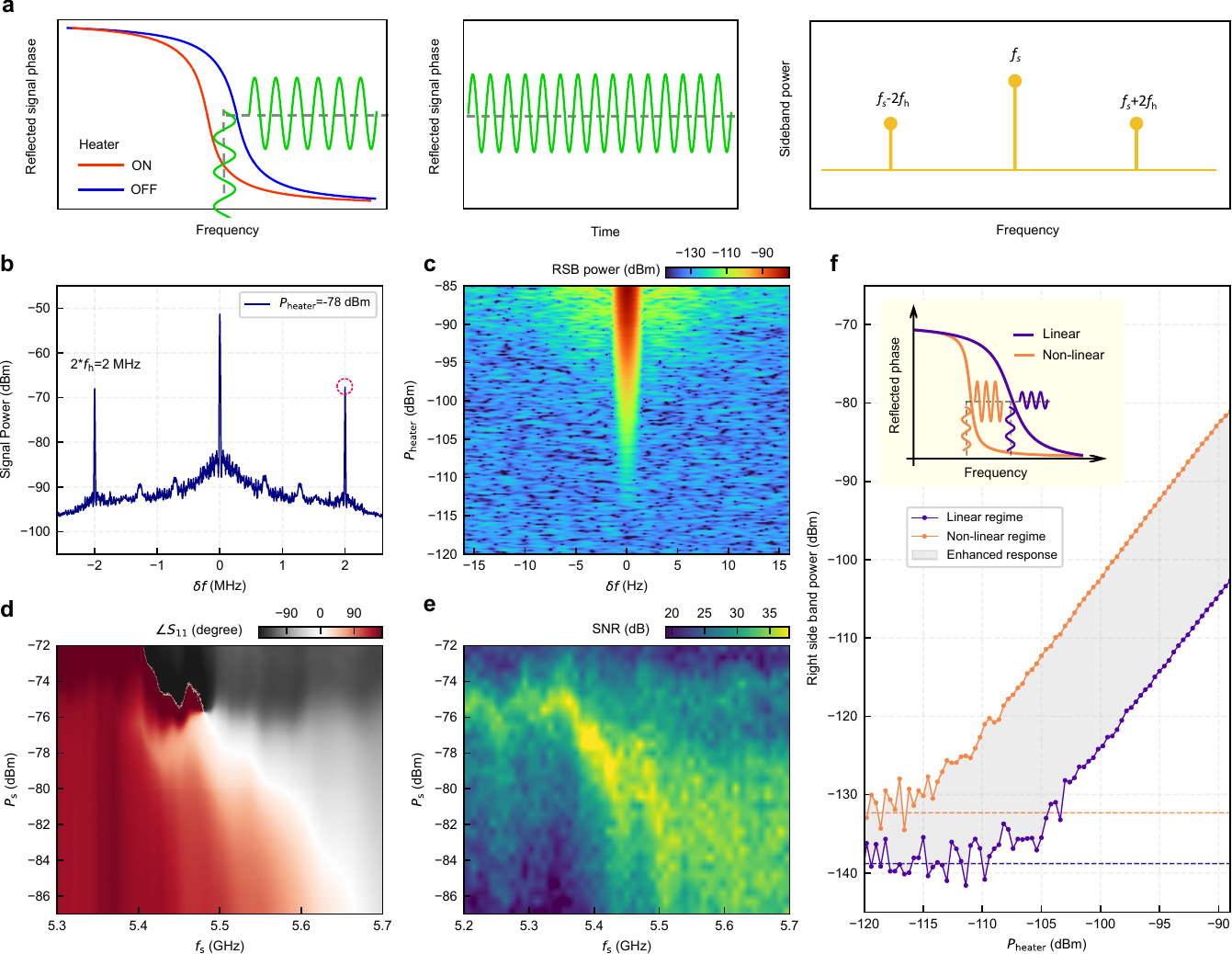}
    \caption{ \label{fig:bolosideband1} { \textbf{Graphene JPA as a sensitive bolometer.}
    \textbf{a,}~Shows the sideband generation scheme. A modulated heater signal causes the phase ($\angle S_{11}$) at a certain frequency to oscillate over time, leading to the generation of sidebands in the frequency domain. The Joule heating effect of any modulated current at frequency $f_{\mathrm{h}}$ produces sidebands at $f_{\mathrm{s}}\pm 2f_{\mathrm{h}}$, where $f_{\mathrm{s}}$ is the pump signal frequency.
    \textbf{b,}~Shows a sideband spectrum of the bolometer device measured on a spectrum analyzer. Here we set the heater signal at $f_{\mathrm{h}}=1$ MHz, which causes the Joule heating in the device and hence modulation of the resonance at $2f_{\mathrm{h}}$. Consequently, we see two sidebands around the pump signal at $f_{\mathrm{s}}\pm 2f_{\mathrm{h}}$.  Here, $\delta f$ is the detuned frequency from the pump signal set at $f_{\mathrm{s}}=5.47$ GHz. The $V_\text{g}$ is set to $6$ V and the heater power $P_\text{heater}=-78$ dBm.
    \textbf{c,}~Shows the sideband signal power measured as a function of heater power ($P_\text{heater}$) and detuned frequency ($\delta f$). Here we zoom on the right sideband as shown by the red dashed circle in Fig.~\ref{fig:bolosideband1}b and then resolve it as a function of $P_\text{heater}$. The x-axis $\delta f$ is the detuned frequency from the right sideband in Fig.~\ref{fig:bolosideband1}b. As we increase the heater power the sideband signal rises and eventually saturates.
    \textbf{d,}~Shows the non-linear phase diagram of the JPA, as a function of signal frequency ($f_{\mathrm{s}}$) and power ($P_{\mathrm{s}}$). The $V_{\mathrm{g}}$ is kept at $6$ V, which fixes the linear resonance at low powers. With increasing signal power the JPA becomes nonlinear and the resonance starts to shift. An increase in power pushes the JPA towards a critical point beyond which any further increase in power makes it unstable (dark grey region). 
    \textbf{e,}~Shows the SNR map of the sideband signal measured as a function of bias signal frequency ($f_{\mathrm{s}}$) and power ($P_{\mathrm{s}}$). Here we keep the $V_\text{g}$ at $6$ V, set the $f_{\mathrm{h}}$ at $1$ MHz, and measure the right sideband signal power at different points of the non-linear phase diagram of Fig.~\ref{fig:bolosideband1}d. We observe that when the JPA bolometer becomes non-linear, there are hotspots of SNR in these regions.
    \textbf{f,}~Shows the sideband power plotted as a function of $P_\text{heater}$ for linear and non-linear biasing of the bolometer at the points (linear: $f_\text{s}=5.68$ GHz, $P_\text{s}=-97$ dBm) and (non-linear: $f_\text{s}=5.39$ GHz, $P_\text{s}=-76.87$ dBm) respectively. In the non-linear regime, the sideband signal is $\sim20$ dB higher, and the sideband starts to rise in $\sim10$ dB lower heater powers than that of the linear regime. This indicates that the device is more sensitive when operating in non-linear biasing. The inset schematic illustrates the phase sensitivity in the non-linear over the linear regime. 
    }}
\end{figure*}

Next, to extract the sensitivity of the bolometer, we perform the heater modulation experiments. In Fig.~\ref{fig:bolosideband1}a, we explain the sideband generation scheme due to modulated heating. A modulated heater signal causes the JPA phase ($\angle S_{11}$) at a certain frequency to oscillate over time, resulting in the generation of sidebands in the frequency domain. The Joule heating effect of any modulated current at frequency $f_{\mathrm{h}}$ produces sidebands at $f_{\mathrm{s}}\pm 2f_{\mathrm{h}}$, where $f_{\mathrm{s}}$ is the pump signal frequency, set near the resonance. Fig.~\ref{fig:bolosideband1}b shows the experimental sideband data from the device measured using a spectrum analyzer, we send a heater signal at a frequency $f_{\mathrm{h}}=1$ MHz and heater power $P_\text{heater}=-78$ dBm. We see two sidebands at twice the frequencies ($2f_{\mathrm{h}}=2$ MHz) along with the pump signal in the middle. The pump signal is set near the non-linear regime of the bolometer with $f_{\mathrm{s}}=5.47$ GHz, $P_{\mathrm{s}}=-79.77$ dBm and $V_{\mathrm{g}}=6$ V. Now, to check the heater sensitivity, in Fig.~\ref{fig:bolosideband1}c we plot the right sideband (RSB) signal power measured as a function of the detuned frequency ($\delta f$) and heater power. As we increase the heater power the sideband signal rises and eventually saturates, indicating saturation of the bolometer.

We quantify the sensitivity of our device through the sideband measurements as discussed in Fig.~\ref{fig:bolosideband1}c. We have observed that when the bolometer is biased in the non-linear regime it is very sensitive to heater modulations. Fig.~\ref{fig:bolosideband1}d shows a non-linear phase diagram of the bolometer, where we measure the reflected phase ($\angle S_{11}$) of the device as a function of microwave signal frequency ($f_{\mathrm{s}}$) and power ($P_{\mathrm{s}}$). This is the typical non-linear phase diagram of a JPA, with a cubic non-linear term in the CPR. Next, at each point of this phase diagram, we bias the bolometer and measure its sideband response for the heater turned on and off. Eventually, we extract a signal-to-noise ratio (SNR) map of the bolometer across the non-linear phase diagram (see Supplementary Fig. S4b,c for details).  Fig.~\ref{fig:bolosideband1}e shows the SNR of the sideband signal measured as a function of bias signal's frequency ($f_{\mathrm{s}}$) and power ($P_{\mathrm{s}}$). Here we keep the gate voltage fixed at $V_\text{g}=6$ V, set the heater frequency at $f_{\mathrm{h}}=1$ MHz, and measure the right sideband signal power. In this measurement, the heater power is kept fixed at $P_\text{heater}=-95$ dBm for the ON state of the bolometer. We observe that when the bolometer becomes non-linear, there are hotspots of high SNR in these regions. These hotspots are preferred for biasing the bolometer.

Next, we bias the device at linear and non-linear regimes to compare its bolometric sensitivity. Fig.~\ref{fig:bolosideband1}f shows the sideband power plotted as a function of heater power $P_\text{heater}$ for linear and non-linear biasing of the bolometer at the biasing points (linear: $f_\text{s}=5.68$ GHz, $P_\text{s}=-97$ dBm) and (non-linear: $f_\text{s}=5.39$ GHz, $P_\text{s}=-76.87$ dBm) respectively. We find in the non-linear regime, the sideband signal is $\sim20$ dB higher, and the signal starts to rise at $\sim10$ dB lower heater powers than that of the linear regime. This indicates that the device is more sensitive when operating in non-linear biasing. Next, we extract the NEP of the device from the data shown in Fig.~\ref{fig:bolosideband1}f, at the non-linear biasing. Supplementary Fig. S5 shows the extracted NEP of the bolometer. The best NEP value for $f_\text{h}=1$ MHz is $\sim500$ aW/$\sqrt{\text{Hz}}$. In graphene bolometer devices, an open question in the field is the extent to which the injected heat couples to the JJ channel. We account for the fact that a finite fraction of the heat injected into the graphene heater leaks to the phonon bath before reaching the JJ. That hot electrons convert to phonons that transfer the heat to hBN is a surprising observation. We experimentally verify this transduction of hot electrons to hBN next.

\begin{figure*}
    \centering
	\includegraphics[width=15cm]{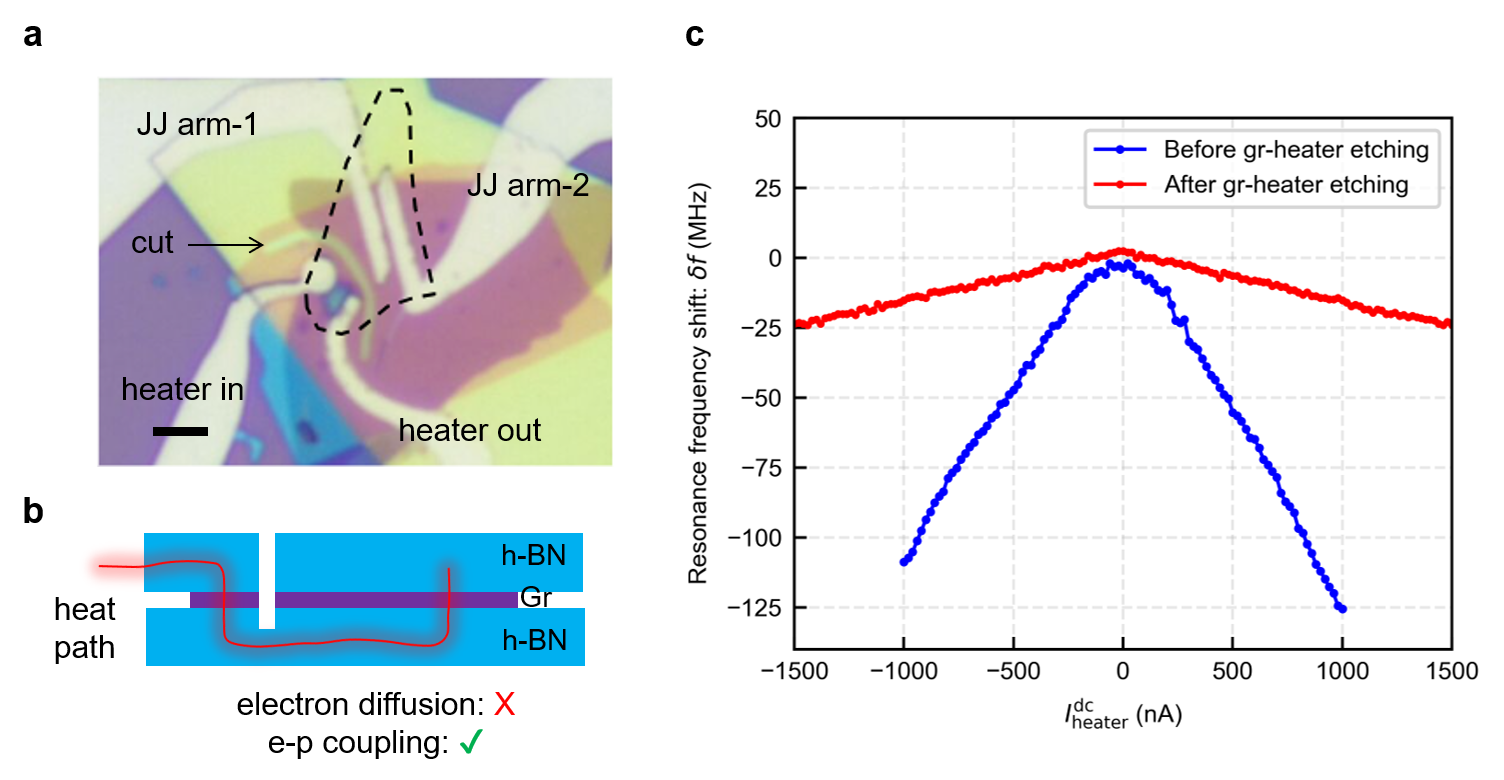}
    \caption{ \label{fig:bolocut} { \textbf{Heat conduction through phonon channels.} 
    \textbf{a,}~Shows an optical image of a control device for testing the heat conduction contribution of electrons and phonons. The JJ and the heater electrode share the same graphene flake marked by a dashed black line. After measuring one round of the device response we etch the graphene heater to create a galvanic discontinuity between the heater and the JJ. The scale bar is $2$ $\micro$m. 
    \textbf{b,}~Shows the cross-sectional schematic of the hBN-gr-hBN stack. The white cut region shows the galvanic discontinuity for electron diffusion. The only possible way for the heat to reach the JJ is through substrate phonons. 
    \textbf{c,}~Shows resonance frequency shift as a function of DC Joule heating as described in Fig.~\ref{fig:boloDCphseshft}a. After etching the graphene the device still gets heated through the substrate phonons and shows a shift in the resonance frequency. 
    }}
\end{figure*}

There are two possible routes for the heat to reach the JJ -- first, the hot electrons diffuse to the junction and second, the phonons excited in the substrate carry heat to the JJ. To test the relative efficacy of these two channels we first study a device with a contiguous graphene connecting the heater and the JJ; subsequently, we sever the galvanic connection by etching the graphene in the same device --  disconnecting the heater and JJ electrically.  After severing the galvanic connection we eliminate the channel of hot electrons pertubing the JJ response while still allowing for the phonons to affect the JJ.  In our control experiment with an etched graphene heater device (see Fig.~\ref{fig:bolocut}a, b), we observe that even after isolating the graphene heater and the JJ by introducing a cut in the graphene, heat can still propagate to the JJ via the phonons in the hBN. This phonon-mediated heat transfer causes a shift in the resonance frequency, as shown in Fig.~\ref{fig:bolocut}c. This indicates that a finite portion of the heat injected into the device leaks to the substrate phonons. Notably, this response is seen despite interfacial Kapitza resistance at multiple interfaces. Studies utilizing scanning probes have demonstrated that hot electrons in graphene dissipate heat to atomic defects through resonant inelastic scattering for cooling~\cite{halbertal_imaging_2017,kong_resonant_2018}. However, such cooling processes at the device level, particularly in configurations involving small JJ channels, remain largely unexplored and challenging~\cite{fried_performance_2024}. Here, our control experiment provides insights into heat dissipation to substrate phonons via atomic defects. Our finite element simulations estimate that approximately $\sim35\%$ of the heat injected at the heater is lost to the phonon bath, as detailed in Supplementary Information, Section XI. See supplementary information section VIII for sideband data at different gate voltages. The on-flake heating scheme in our device works from DC to $100$ MHz heater signal frequencies; see supplementary information section VII for sideband data at different heater frequencies.

\begin{figure*}
    \centering
	\includegraphics[width=14cm]{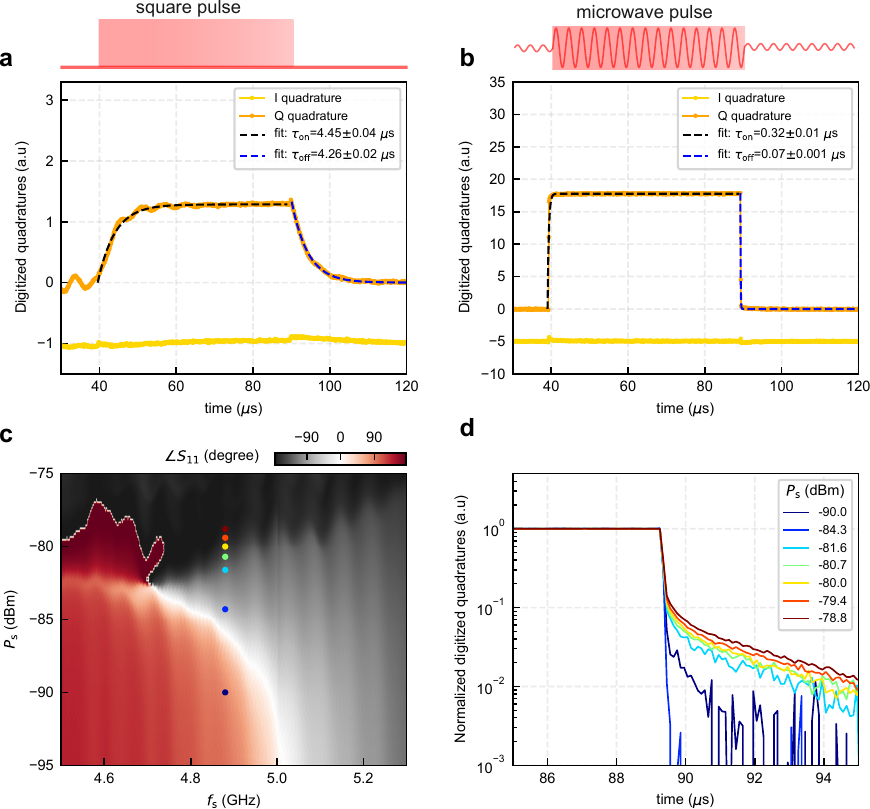}
    \caption{ \label{fig:bolothermT} { \textbf{Time constants of the bolometer.} 
    \textbf{a,}~Shows the measured thermal time constant of the bolometer. We heat the bolometer using a short DC square pulse and simultaneously probe its response with a pump signal at $4.848$ GHz, in the linear biasing of the device. The heater pulse width is set at $50$ $\mu$s. We measure the digitized quadrature signal of the pump tone using a homodyne demodulation scheme. The quadrature signals ($I, Q$) are plotted as a function of time. An exponential fit to the rising and falling edge yields the thermal time constants $\tau_\text{on}=4.45$ $\mu$s and $\tau_\text{off}=4.26$ $\mu$s. 
    \textbf{b,}~Shows the measured intrinsic time constant of the JPA. We probe the JPA using an amplitude-modulated microwave pulse, in the linear biasing of the device. The microwave pulse is made using a pump signal of $4.88$ GHz mixed with a square pulse for amplitude modulation. The square pulse width is set at $50$ $\mu$s. We measure the digitized quadrature signal of the pump pulse using a homodyne demodulation scheme. The quadrature signals ($I, Q$) are plotted as a function of time. An exponential fit to the rising and falling edge yields the intrinsic JPA time constants $\tau_\text{on}=0.32$ $\mu$s and $\tau_\text{off}=0.07$ $\mu$s. 
    \textbf{c,}~Shows the non-linear phase diagram of the JPA plotted as a function of signal frequency ($f_{\mathrm{s}}$) and power ($P_{\mathrm{s}}$). The gate voltage is kept at $V_{\mathrm{g}}=2$ V.
    \textbf{d,}~Shows the measured intrinsic time constant of the JPA, for different pump signal powers marked by colored dots in the non-linear phase diagram of Fig.~\ref{fig:bolothermT}c. We probe the JPA using amplitude-modulated microwave pulses. We measure the digitized quadrature signal of the pump pulse using a homodyne demodulation scheme. The quadrature signals ($I, Q$) are plotted as a function of time. For small signal powers the device responds fast, however for large signal powers the device becomes non-linear and eventually dissipative; hence, the response time lags and develops multiple time constants,  evident from the slope change in the log scale.   
    }}
\end{figure*}

Next, we measure the time constant of the bolometer experimentally. The response time of a bolometer quantifies the time required to excite and reset a bolometer. Using the graphene heater port, we heat the bolometer using short DC pulses sent in sequences while simultaneously probing its response with a pump signal at the resonance frequency of $4.848$ GHz. The heater pulse width is set at $50$ $\mu$s. We measure the digitized quadrature signal of the pump tone using a homodyne demodulation scheme (see supplementary information section VI for setup details). The quadrature signals ($I, Q$) are plotted as a function of time, see Fig.~\ref{fig:bolothermT}a. An exponential fit to the rising and falling edge yields the thermal time constants $\tau_\text{on}=4.45$ $\mu$s and $\tau_\text{off}=4.26$ $\mu$s. The data is taken at a fixed gate voltage and measured with averaging over multiple pulse sequences. The device shows a fast response time, comparable to previous studies on graphene JJ-based bolometers~\cite{kokkoniemi_bolometer_2020}. Different cooling channels can contribute to thermalizing the hot carriers in graphene such as phonons and electrons. In graphene JJs electron diffusion is suppressed due to Andreev reflection of the carriers at the superconducting contacts and suppresses the cooling through diffusion. Hence, phonons play an important role in thermalizing the graphene's electrons to the bath. However, at dilution fridge temperatures, $\sim20$ mK, phonon occupancies are small and can explain the relatively slow thermal time constant of the bolometer. 

Next, to verify that the bolometer's time response is not constrained by the intrinsic time scale of the JPA, we measure the time response of the JPA. We drive the JPA with an amplitude-modulated microwave pulse and simultaneously probe it through the digitized quadrature signals using a homodyne demodulation scheme (see supplementary information section VI for setup details). During this experiment, the heater is not energized. At low powers, the JPA exhibits a faster response with $\tau_\text{on}=320$ ns and $\tau_\text{off}=70$ ns, as shown in Fig.~\ref{fig:bolothermT}b. The response time of the JPA is limited by the intrinsic bandwidth of the device. Next, we vary the signal power of the microwave pulse to drive the JPA from the linear regime to the non-linear regime and eventually to the normal state of the JJ, following the probing points marked by the colored dots in Fig.~\ref{fig:bolothermT}c. In the linear regime, the JPA responds quickly, but as the power increases, the response slows down, as shown in Fig.~\ref{fig:bolothermT}d. This increase in response time at high power is expected in the non-linear regime of the JPA, where the internal bandwidth of the device decreases and also due to local heating at the JJ.

In summary, we report the experimental realization of a bolometer device using a graphene JPA for the first time. The Kerr non-linearity of the JPA helps in enhancing the bolometer sensitivity. When the bolometer is biased in the non-linear regime, it enhances the sideband signals ($\sim100$ times), resulting in an order of magnitude improvement in sensitivity compared to the linear regime. We achieve a best NEP~$\sim500$ aW/$\sqrt{{\mathrm{Hz}}}$. We experimentally measure a fast thermal time constant of the bolometer of $4.26$ $\mu$s. The large operating bandwidth and fast measurement scheme are key features of our device that will improve upon previous sensing schemes that relied on shifts in switching histograms to demonstrate exquisite sensitivity. Our device has the potential to sense THz and NIR photons via a direct irradiation mechanism, where the graphene flake acts as a bus for carrying the hot electrons, leading to a frequency shift in the device. The sensitivity of the bolometer can be improved further by using superconductors with a smaller energy gap, such as aluminum, for fabricating the JJs.

Our work extends the pathway for the exploration of $2$D van der Waals materials-based devices for future-generation quantum sensors. Fast broadband bolometers have versatile applications in readout processes for cQED experiments~\cite{opremcak_measurement_2018,gunyho_single-shot_2024}, suggesting that our device could also be utilized for similar measurements. Recent experiments on MKIDs and SNSPDs have demonstrated significant pixel multiplexing capabilities in bolometer devices~\cite{walter_mkid_2020,oripov_superconducting_2023}, which could also be adapted in large-scale graphene JPA bolometer devices for multiplexing applications. Furthermore, the nonlinear sensitivity enhancement, a novel and critical feature unutilized in MKIDs and SNSPDs, is feasible within our device architecture, while still enabling fast readout capabilities similar to those of MKIDs and SNSPDs.

\section*{Acknowledgements:}
We thank Vibhor Singh, Eli Zeldov, Vishal Ranjan, and R. Vijay for their helpful discussions and comments. We thank Kishor V. Salunkhe, Mahesh Hingankar, and Digambar A. Jangade for their experimental assistance. We acknowledge the Nanomission grant SR/NM/NS-45/2016 and DST SUPRA SPR/2019/001247 grant along with the Department of Atomic Energy of Government of India 12-R\&D-TFR-5.10-0100 for support. We acknowledge AOARD grant FA2386-23-1-4031 for support.
Preparation of hBN single crystals is supported by the Elemental Strategy Initiative conducted by the MEXT, Japan (Grant Number JPMXP0112101001) and  JSPS KAKENHI (Grant Numbers 19H05790 and JP20H00354).

\section*{Author Contributions:}
J.S. fabricated the devices, led the measurements, and analyzed the data. K.M. and P.S. helped with measurements and analysis. A.S., H.A., and P.S. assisted in the device fabrication. R.R. and A.S. assisted in the finite element simulations. A.B. assisted in the time constant measurements. M.P.P. assisted in microwave PCB fabrications. K.W. and T.T. grew the hBN crystals. J.S. and M.M.D. wrote the manuscript with input from everyone. M.M.D. supervised the project.

\clearpage

\clearpage

\begin{center}
\textbf{\huge Supplementary Information}
\end{center}
\renewcommand{\thesection}{\Roman{section}}
\setcounter{section}{0}
\renewcommand{\thefigure}{S\arabic{figure}}
\captionsetup[figure]{labelfont={bf},name={Supplementary Fig.}}
\setcounter{figure}{0}
\renewcommand{\theHequation}{Sequation.\theequation}
\renewcommand{\theequation}{S\arabic{equation}}
\setcounter{equation}{0}

\section{DC characterization of the bolometer device}\label{sec:boloDCcheck}
\begin{figure}[H]
    \centering
	\includegraphics[width=15cm]{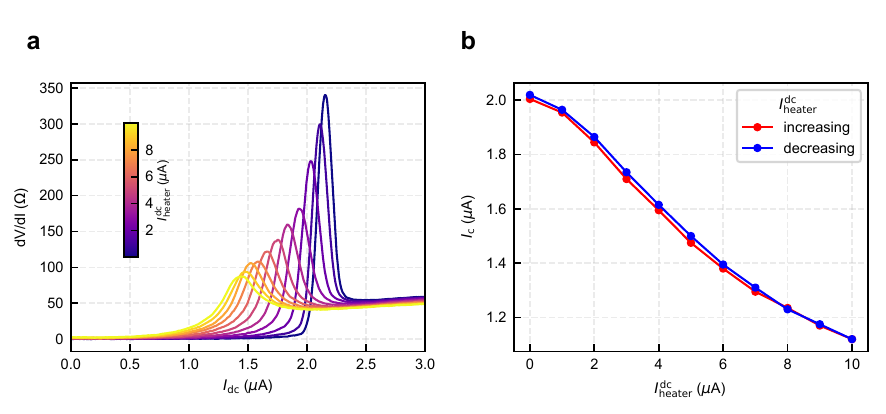}
    \caption{\label{fig:boloDCchar} {\textbf{DC characterization of the bolometer.} 
        \textbf{a,}~Shows a pseudo-$4$-probe differential resistance ($\mathrm{d}V/\mathrm{d}I$) of the graphene-JJ as a function of dc-bias current ($I_{\mathrm{dc}}$) for increasing heater currents ($I_{\mathrm{heater}}^{\mathrm{dc}}$). The heater current is sent from the heater port and drains to the ground. The extended part of the graphene, which is away from the junction, is resistive. As a result, the Joule heating caused by the heater current leads to a local increase in the junction temperature. Hence, the switching current ($I_{\text{c}}$) decreases and the transition peaks also broaden. 
	\textbf{b,}~Shows the switching current ($I_{\text{c}}$) of the graphene-JJ as a function of dc heater current ($I_{\mathrm{heater}}^{\mathrm{dc}}$) for increasing and decreasing directions of the heater current. Both curves fall on top of each other indicating good contact of the heater and minimal hysteresis in the device. 
    }}
\end{figure}

The DC measurements primarily involve the pseudo-$4$-probe differential resistance measurements of the JJs using a standard lock-in detection technique with an SRS 830 lock-in amplifier. We send small ($\sim50$ nA$\ll I_{\mathrm{c}}$) low frequency ($17$ Hz) current through the parallel LC resonator and measure the voltage across it as a function of DC bias current ($I_{\mathrm{dc}}$) and applied back gate voltage ($V_{\mathrm{g}}$); due to the frequency of the AC current being very small the capacitor can effectively be thought as open. Hence the voltage across the parallel LC we assume to be the response of the JJ only. We take precautions to filter any undesired high-frequency signals that may travel from room temperature to the device. Three-stage filtering of the DC lines is done using low pass RC filters that are kept at different temperature plates of our dilution fridge (room temperature, $4$ K plate, $20$ mK plate) and they all have cut-off frequencies $\sim70$ kHz. The line used for the back gate ($V_{\mathrm{g}}$) also passes through the aforementioned three-stage RC filtering; additionally, we put a $10$ Hz low pass RC filter at room temperature. In addition to the RC filters we also have copper powder filters (at $20$ mK) and eccosorb filters (at $4$ K) in our measurement lines to attenuate any higher frequency noise.\\
We measure the differential resistance ($\mathrm{d}V/\mathrm{d}I$) of the graphene-JJ as a function of dc-bias current ($I_{\mathrm{dc}}$) for increasing heater currents ($I_{\mathrm{heater}}^{\mathrm{dc}}$), shown in Fig.~\ref{fig:boloDCchar}. As a result of the Joule heating caused by the heater current, the local temperature of the junction increases. Hence, the switching current ($I_{\mathrm{c}}$) decreases and the transition peaks also broaden. 

\section{Microwave measurements of the bolometer device}\label{sec:boloRFsetup}
\begin{figure}[H]
	\centering
	\includegraphics[width=13cm]{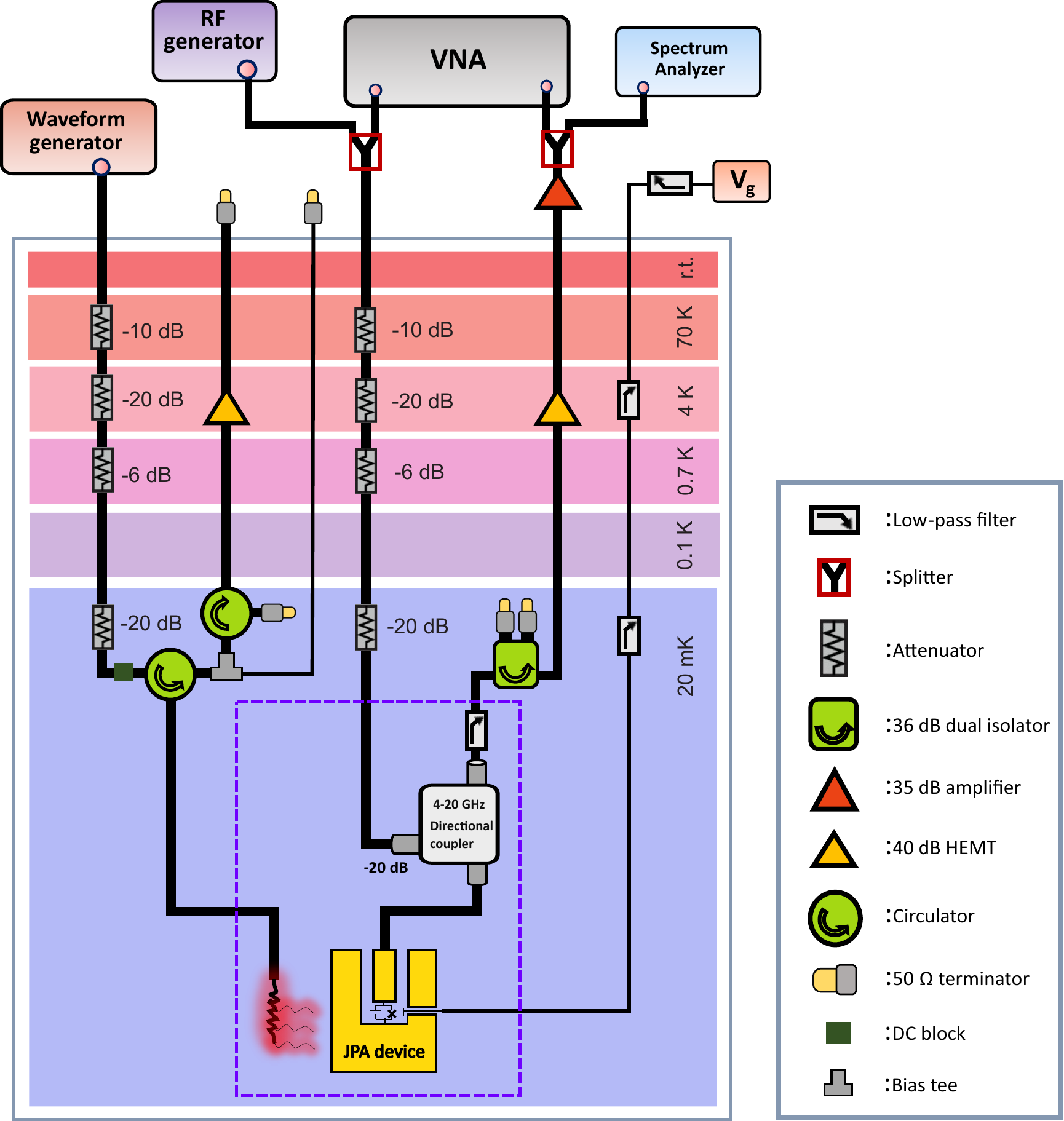}
	\caption[Wiring diagram and the microwave measurements setup]{ \label{fig:bolosetupwiring} {\textbf{Wiring diagram and the microwave measurements setup.}}
		{Shows the wiring diagram for the bolometer measurements. The blue dashed box indicates the components that are inside the puck. We insert the puck through a bottom load locking system in our dilution fridge. The various microwave components are listed in the box. A separate microwave line, which is already present inside our fridge for reflection measurements, is used for the heater line. 
	}}
\end{figure}

The microwave measurements involve the standard reflectometry technique. Full characterization of the bolometer involves checking its gate tunability, nonlinear response, sideband measurements, SNR, and thermal time constant measurements. Similar to DC setup in microwave measurements we take precautions for filtering any undesired noise that may reach the device. The microwave setup along with the wiring diagram is shown in Fig.~\ref{fig:bolosetupwiring}. The microwave lines are made up of stainless steel rigid coaxial lines. The input lines are attenuated by $56$ dB using fixed attenuators at different temperature stages. The output signal from the bolometer passes through a directional coupler (Krytar 4-20 GHz), a 10 GHz low-pass filter\footnote{We have found that this low-pass filter helps in blocking the Johnson noise of the $4$ K HEMT back streaming to the JPA.}, and a dual isolator (LNF) at the $20$ mK plate. The output signal is amplified using a $40$ dB high-electron-mobility transistor (HEMT) amplifier at the $4$ K plate, and a $35$ dB room temperature amplifier. We use a VNA (R\&S ZNB 20) for basic reflection measurements of the device. We use a separate RF source (SignalCore RF generator) to send the continuous wave (CW) pump tone. We use a spectrum analyzer (R\&S FSV signal analyzer) for analyzing output signal powers. The gate voltage ($V_{\mathrm{g}}$) is applied to the bolometer using a DC voltage source (NI-DAQ). The DC line for the gate passes through a $10$ Hz low pass RC filter followed by three-stage filtering using low pass RC filters that are kept at different temperature plates of our dilution fridge (room temperature, $4$ K plate, $20$ mK plate) and they all have cut-off frequencies $\sim70$ kHz. A separate microwave line, which is already present inside our fridge for reflection measurements, is used for the heater line.  We use an arbitrary waveform generator (AWG, Agilent 33220A) for sending the heater signal. The bolometer device is loaded inside a copper PCB box and an aluminum puck for cryogenic electromagnetic shielding. The thermal time constant measurement setup is discussed in the next subsection.

\section{Heater modulation and sideband generation scheme}
A modulated heater signal will cause the resonator phase ($\angle S_{11}$) at a certain frequency to oscillate over time, resulting in the generation of sidebands in the frequency domain. The heating effect of any modulated heater current at frequency $f_{\mathrm{h}}$ produces sidebands at $f_{\mathrm{s}}\pm 2f_{\mathrm{h}}$, where $f_{\mathrm{s}}$ is the pump signal frequency, parked near the resonance. We can measure the phase oscillation with time at low-frequency heater modulations probed using a VNA, discussed in supplementary section IX. 

\begin{figure}[H]
	\centering
	\includegraphics[width=9cm]{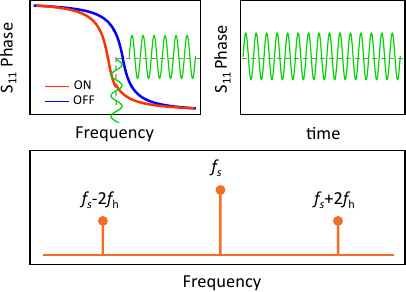}
	\caption[Heater modulation and sideband generation scheme]{ \label{fig:bolosideband} {\textbf{Heater modulation and sideband generation scheme.}}
		{Shows the sideband generation scheme due to modulation of the heater current. A modulated heater signal will cause the bolometer phase ($\angle S_{11}$) at a certain frequency to oscillate over time, resulting in the generation of sidebands in the frequency domain. The heating effect of any modulated heater current at frequency $f_{\mathrm{h}}$ produces sidebands at $f_{\mathrm{s}}\pm 2f_{\mathrm{h}}$, where $f_{\mathrm{s}}$ is the pump signal frequency, parked near resonance.
	}}
\end{figure}

\section{SNR measurement of the bolometer}
Here, we will discuss the algorithm for measuring the SNR diagram of the bolometer. For a given heater frequency, there are two sidebands at \( f_s \pm 2f_h \), where \( f_s \) is the pump signal frequency and \( f_h \) is the heater frequency. Our goal is to measure the right sideband power at the frequency \( f_s + f_h \) using the spectrum analyzer in zero-span continuous wave (CW) mode. We will refer to the measured right sideband power as \( P_{\mathrm{RSB}} \). By varying the pump signal frequency \( f_s \) and the pump signal power \( P_s \), we will create a 2D map. Thus, we define the SNR as follows:
\begin{equation*}
    \mathrm{SNR}=\frac{P_{\mathrm{RSB}}^{\mathrm{on}}(f_s,~P_s)}{P_{\mathrm{RSB}}^{\mathrm{off}}(f_s,~P_s)}
\end{equation*}
where "on/off" indicates whether the heater source is active. Now, let’s discuss the instruments involved: a Keysight AWG generates a continuous tone for the heater at a fixed power and frequency \( f_h \). A SignalCore RF generator then sweeps the signal frequency \( f_s \) and power \( P_s \), with power as the fast axis and frequency as the slow axis. For each point in the \((f_s, P_s)\) space, an R\&S spectrum analyzer measures the signal power at \( f_s + f_h \) in CW mode, using a $1$ Hz RBW, VBW, and averaging between $100$ to $200$ samples. We first collect the heater ON data, then the heater OFF data, and finally extract the SNR by subtracting the matrices. 

\begin{figure}[H]
    \centering
	\includegraphics[width=15cm]{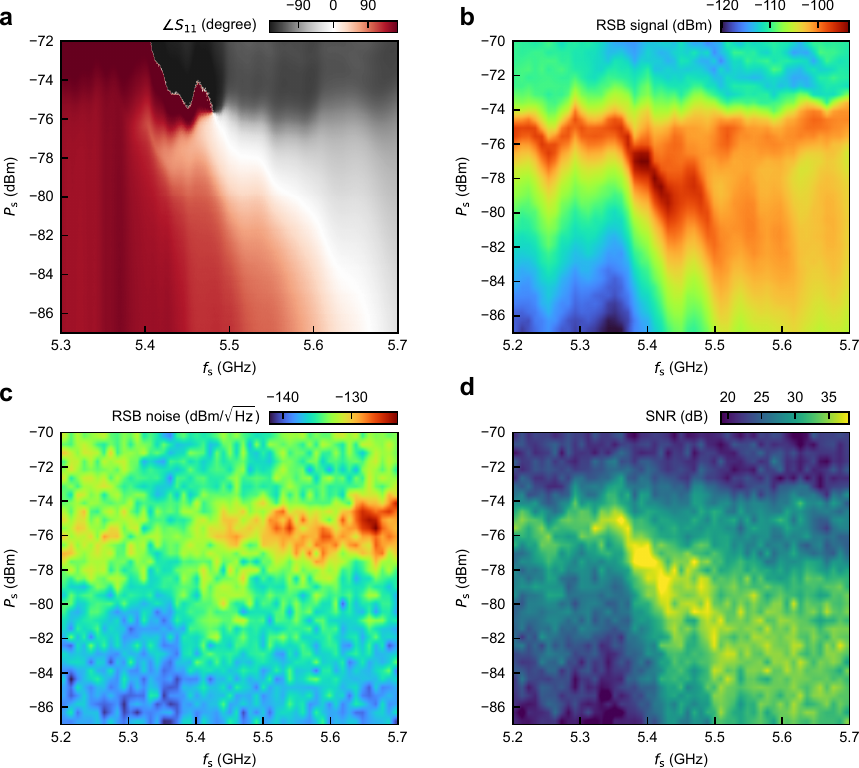}
    \caption{ \label{fig:bolononlinSNR} { \textbf{Graphene JPA as a sensitive bolometer.} 
    \textbf{a,}~Shows the reflected phase ($\angle S_{11}$) of the resonator plotted as a function of microwave signal frequency ($f_{\mathrm{s}}$) and power ($P_{\mathrm{s}}$). Here we fix the gate voltage at $V_{\mathrm{g}}=6$ V, which fixes the linear resonance (\flin) at low powers. With increasing signal power the resonator becomes nonlinear and the resonance starts to shift towards the left side of linear resonance. An increase in power pushes the resonator towards a critical point beyond which any further increase in power makes it bistable and then unstable (dark grey region). 
    \textbf{b,}~Shows the signal power of the sideband signal measured as a function of biasing signal frequency ($f_{\mathrm{s}}$) and power ($P_{\mathrm{s}}$). Here we keep the gate voltage fixed at $V_{\mathrm{g}}=6$ V, set the heater frequency at $f_{\mathrm{h}}=1$ MHz, and measure the right sideband signal power at different points of the non-linear phase diagram of Fig.~\ref{fig:bolononlinSNR}a. In this measurement, the heater power is kept at $P_{\mathrm{heater}}=-95$ dBm. Near the non-linear regime the sideband signal increases.
    \textbf{c,}~Shows the noise power measured as a function of biasing signal frequency ($f_{\mathrm{s}}$) and power ($P_{\mathrm{s}}$). Here we keep the gate voltage fixed at $V_{\mathrm{g}}=6$ V, set the heater frequency at $f_{\mathrm{h}}=1$ MHz, and measure the noise power at different points of the non-linear phase diagram of Fig.~\ref{fig:bolononlinSNR}a. In this measurement, the heater is kept OFF.  
    \textbf{d,}~Shows the signal-to-noise ratio (SNR) of the sideband signal measured as a function of biasing signal frequency ($f_{\mathrm{s}}$) and power ($P_{\mathrm{s}}$). Using Fig.~\ref{fig:bolononlinSNR}b,c we construct the 2D map of the SNR diagram. We observe that when the JPA bolometer becomes non-linear, there are hotspots of SNR in these regions. These hotspots are preferred for biasing the JPA bolometer.
    }}
\end{figure}

\section{NEP extraction}
The Noise Equivalent Power (NEP) of a thermal detector quantifies the minimum power it is capable of sensing. Consider a power-to-voltage converter has a responsivity ($R$), such that $V_{\text{out}} = R(P_{\text{in}})$, where $V_{\text{out}}$ is the output voltage signal and $P_{\text{in}}$ is the applied input power. In the linear-response regime, for small applied power, this expression can be written as:
\begin{equation*}
    \delta V_{\text{out}} \approx \left( \frac{\partial V_{\text{out}}}{\partial P_{\text{in}}} \right) \ast \delta P_{\text{in}}.
\end{equation*} 
In this regime, the NEP of a power-to-voltage detector is defined as: $ NEP = \sqrt{S_v}/\left( \frac{\partial V_{\text{out}}}{\partial P_{\text{in}}} \right)$ where, \( \sqrt{S_v} \) is the measured voltage spectral density and \( \frac{\partial V_{\text{out}}}{\partial P_{\text{in}}} \) is the device responsivity. Hence, NEP can be extracted by measuring the voltage spectral density \( \sqrt{S_v} \) at the output and the device responsivity \( \frac{\partial V_{\text{out}}}{\partial P_{\text{in}}} \). Also, by measuring the applied power at the input \( \delta P_{\text{in}} \), and the SNR at the output, as indicated by rearranging the above expression\footnote{We use this expression for measuring the NEP, where we measure the heater power in units of W and SNR in units of V/(V/$\sqrt{\text{Hz}}$), which gives NEP in units of W/$\sqrt{\text{Hz}}$.}

\begin{equation}\label{eqn:boloSNEPexprssn}
    NEP = \frac{\sqrt{S_v}}{\delta V_{\text{out}}} \ast \delta P_{\text{in}} = \frac{\delta P_{\text{in}}}{SNR}.
\end{equation}
We account for the fact that a finite fraction of the heat injected into the graphene heater leaks to the phonon bath before reaching the JJ. Also, we take into account the impedance mismatch of the heater resistance ($\sim500~\Omega$) from $50~\Omega$ line impedance. The modified expression of the heat reaching the JJ is given by:
\begin{equation}\label{eqn:boloSNEPexprssn}
    P_{\mathrm{heater}}^{~\prime}=P_{\mathrm{heater}}*(1-|\Gamma|^2)*(1-\eta_{e-p}).
\end{equation}
Where $P_{\mathrm{heater}}$ represents the heater power applied to the sample, estimated based on the power set at the instrument level and the fixed line attenuators. The $\Gamma=\frac{R_{\mathrm{heater}}-50}{R_{\mathrm{heater}}+50}$ is the signal reflection due to the impedance mismatch of the heater and $\eta_{e-p}$ is the fraction of the heater power lost at substrate phonon.
\begin{figure}[H]
	\centering
	\includegraphics[width=9cm]{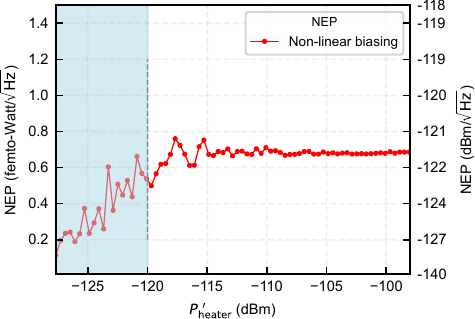}
	\caption[NEP extraction]{ \label{fig:boloNEP} {\textbf{NEP extraction.}}
		{Shows the extracted NEP of the bolometer plotted as a function of heater power $P_{\mathrm{heater}}^{~\prime}$. The NEP is extracted from the non-linear sideband data shown in Main Fig.3f in the non-linear regime, using the prescription given by Eq.~\eqref{eqn:boloSNEPexprssn}. The best NEP value for $f_{\mathrm{h}}=1$ MHz is $\sim500$ aW/$\sqrt{\text{Hz}}$. The error bar represents the proportional error in the NEP extraction. The blue-shaded region corresponds to the unresolved sideband signal buried under the noise floor. 
	}}
\end{figure}

\section{Thermal time constant measurements of the bolometer}\label{sec:boloThermalTschem}
\begin{figure}[H]
	\centering
	\includegraphics[width=12cm]{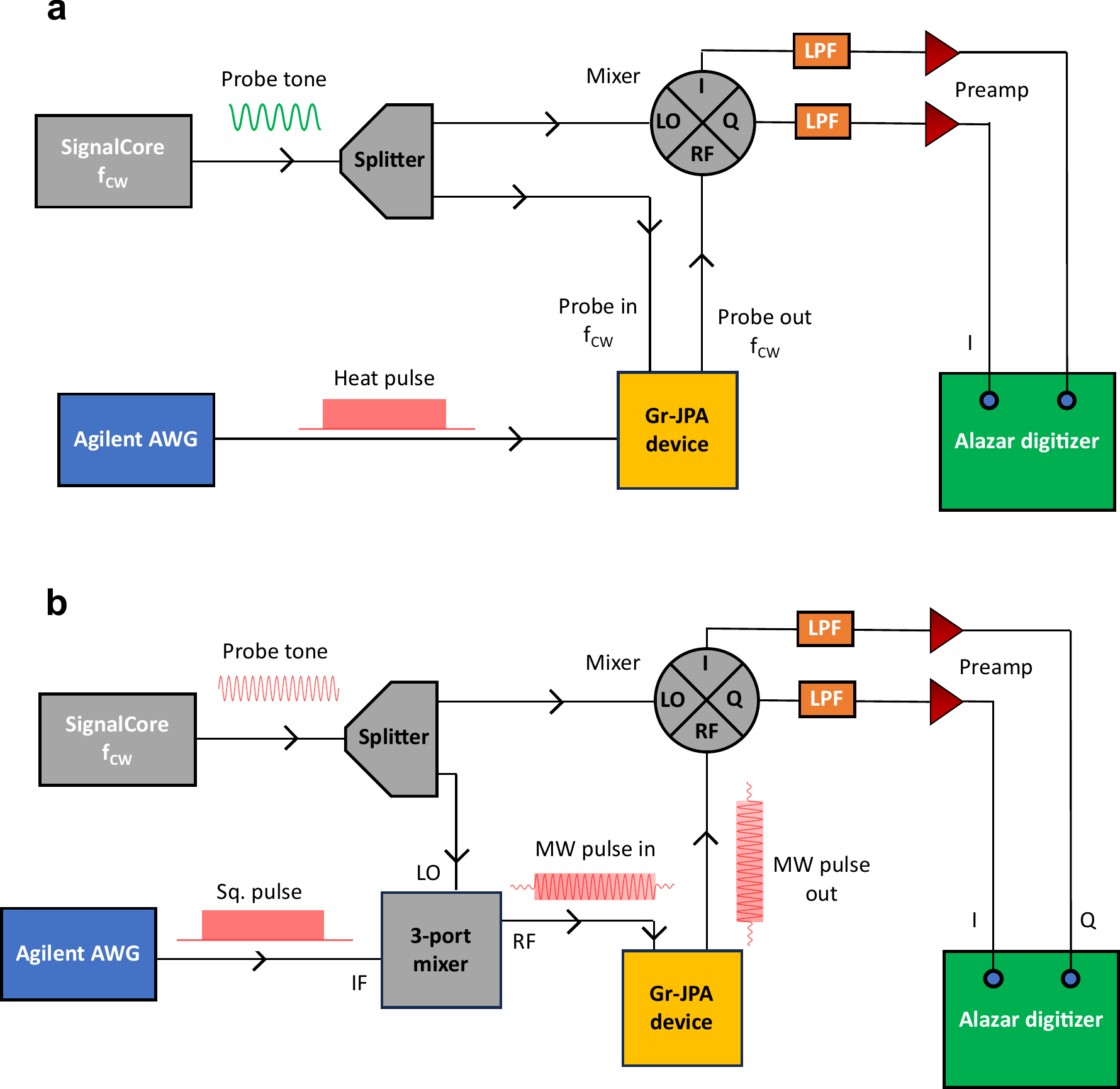}
	\caption[Thermal time constant measurement scheme]{ \label{fig:bolothermTschm} {\textbf{Thermal time constant measurement scheme.}}
		{
			\textbf{a,b}~Shows a standard homodyne demodulation technique, where a pump tone is split, with half sent to the LO port of a $4$-port mixer, and the remaining portion directed to the bolometer after putting desired attenuation. The reflected signal from the bolometer is fed to the RF port of the mixer. The bolometer is heated by sending square pulses sequentially. The demodulated I/Q signals are filtered using low-pass filters, then amplified, and digitized using a fast ADC card at a sampling rate of $10$ MS/s. 
	}}
\end{figure}

We employ a standard homodyne demodulation technique to measure the bolometer's thermal time constant. Here, we probe the bolometer with a continuous wave microwave signal and heat it by sending square pulses. The reflected signal from the bolometer is then demodulated using a 4-port mixer and digitized using a fast ADC card. The components used for this experiment are: Keysight 33600A AWG, Marki MMIQ0416L $4$ port mixer, SignalCore SC5511A RF generator, Alazar ATS9870 digitizer, SR445A amplifier (5x followed by 5x) and minicircuits low pass filters 11 MHz LPF. The schematic of the demodulation setup is shown in Fig.~\ref{fig:bolothermTschm}a. We also measure the JPA time response by driving it with amplitude-modulated microwave pulses. The measurement scheme is shown in Fig.~\ref{fig:bolothermTschm}b.A 3-port mixer (Mini-Circuits ZMX-10G+) is used to generate the amplitude-modulated microwave pulse that drives the JPA from the linear to the non-linear regime by varying the power. The time response of the quadrature signals is then analyzed to study this transition. The rest of the components are similar to that of the first setup. The experimental data for the time constants of the bolometer is shown in the main manuscript Fig. 4.

\vspace{0.3cm}

\section{Bolometer response at different heater frequencies}\label{sec:bolosidebdiffreq}
\begin{figure}[H]
	\centering
	\includegraphics[width=15cm]{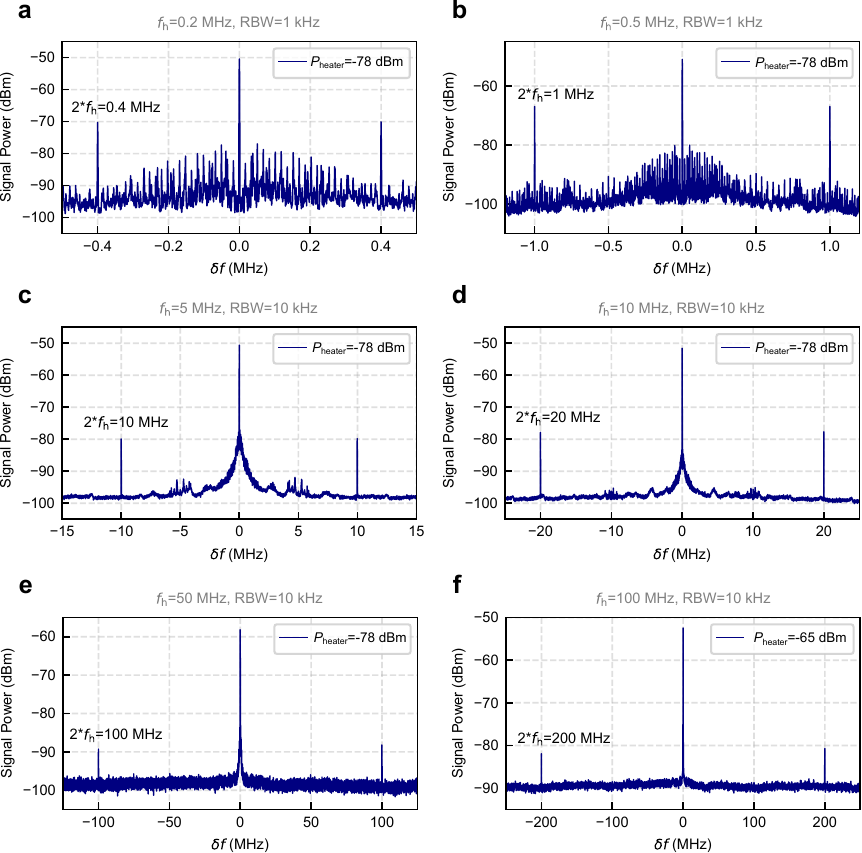}
	\caption[Sideband measurements at higher heater frequencies]{ \label{fig:bolofh0p2} {\textbf{Sideband measurements at higher heater frequencies.}}
		{
			\textbf{a-f}~Shows sideband measurements of the bolometer across various heater frequencies ($f_{\mathrm{h}}$). The heater frequency and resolution bandwidth of the spectrum analyzer are indicated in the titles of each plot. The heater power ($P_{\mathrm{heater}}$) is kept constant.
	}}
\end{figure}

\begin{figure}[H]
	\centering
	\includegraphics[width=9cm]{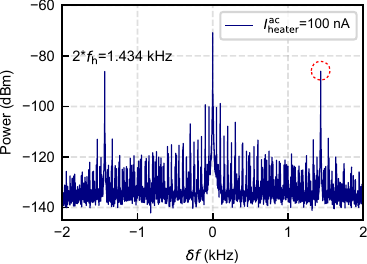}
	\caption[Sideband measurement at low heater frequency]{ \label{fig:bololowfreqsideb} {\textbf{Sideband measurement at low heater frequency.}}
		{
			Shows the measured sideband spectrum of the bolometer. Here we put an AC sinusoid of frequency $f_{\mathrm{h}}=717$ Hz on the heater which causes the heating in the device and hence modulation of the resonance at twice the frequency $2f_{\mathrm{h}}$. Consequently, we see two sidebands at frequencies $f_{\mathrm{s}}\pm 2f_{\mathrm{h}}$ around the pump tone.  In this plot, $\delta f$ is the detuned frequency from the pump tone.
	}}
\end{figure}

\begin{figure}[H]
	\centering
	\includegraphics[width=9cm]{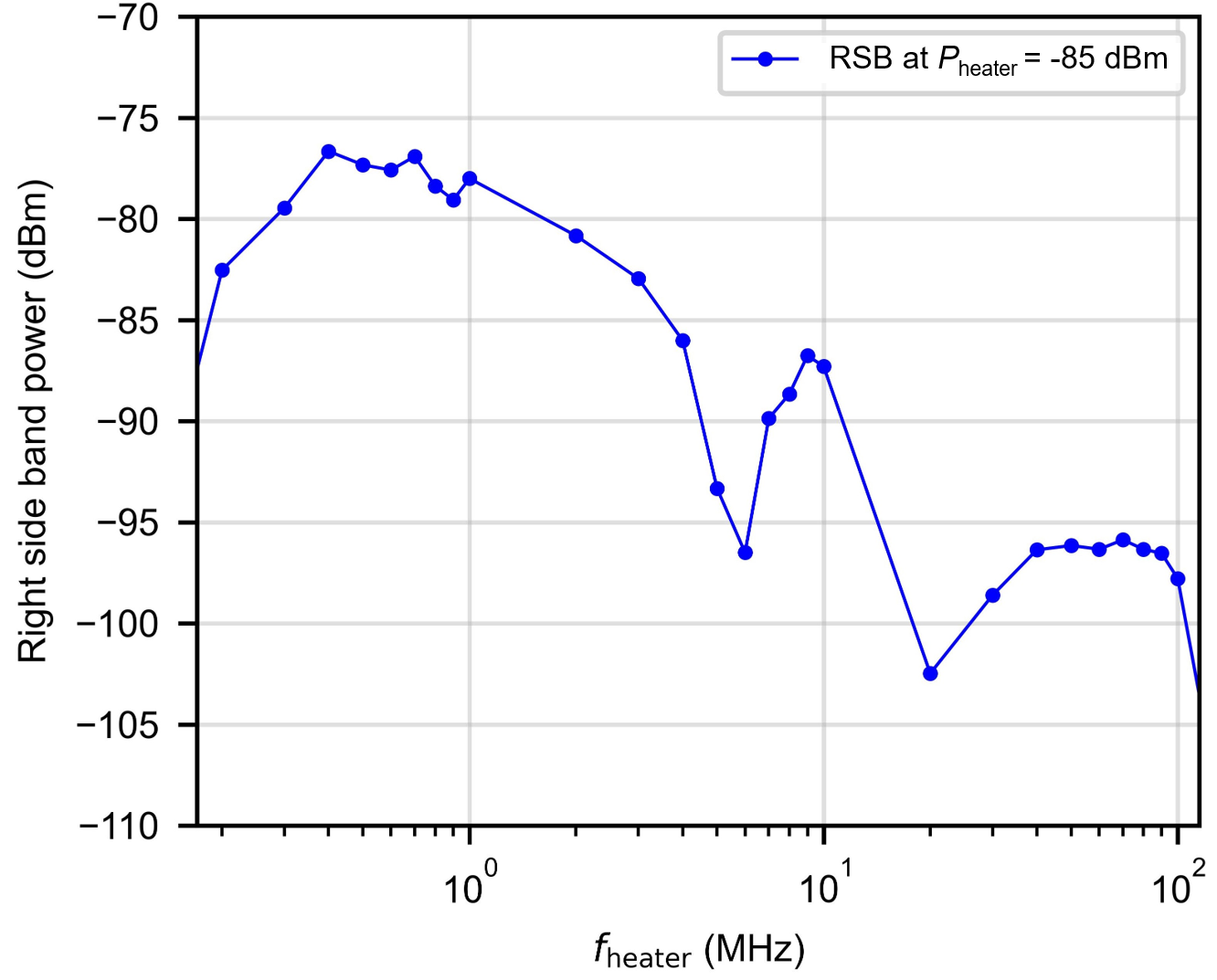}
	\caption[Sideband signal as a function of heater frequency]{ \label{fig:bolosidebndfh} {\textbf{Sideband signal as a function of heater frequency.}}
		{
			Shows the sideband signal of the bolometer plotted as a function of heater frequency. The heater power ($P_{\mathrm{heater}}$) is kept constant. The sideband signal drops at larger frequencies due to impedance mismatches or, heat accumulation effects.
	}}
\end{figure}

Here, we discuss the bolometer response for different heater frequencies ($f_{\mathrm{h}}$), see Fig.~\ref{fig:bololowfreqsideb} and ~\ref{fig:bolofh0p2}. At low very low frequencies (sub $10$ kHz) the bolometer picks up noise from $50$ Hz ground loop interference, leading to noisy sideband response. However, at higher heater frequencies such effects get suppressed. In fig.~\ref{fig:bolosidebndfh} we plot the sideband signal power as a function of heater frequency for a fixed heater power. We see that the sideband signal drops at higher frequencies due to impedance mismatches or, heat accumulation. Further studies are needed to examine this carefully and to make possible optimizations.

\vspace{0.3cm}

\vspace{0.3cm}

\section{Bolometer response at different gate voltages}\label{sec:bolosiddiffeVg}
\begin{figure}[H]
	\centering
	\includegraphics[width=9cm]{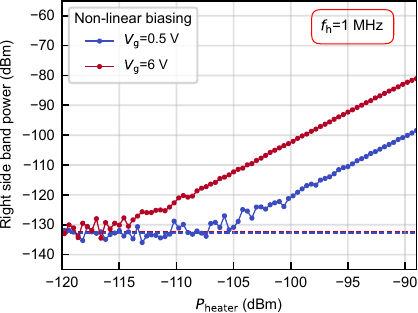}
	\caption[Sideband signal at different gate voltages]{ \label{fig:bolosidebndVg} {\textbf{Sideband signal at different gate voltages.}}
		{
			Shows the sideband signal of the bolometer at different gate voltages. We park the device in the non-linear regime where the SNR is maximum, similar to the method we discussed in Main Fig.3d, and measure the sideband response for different gate voltages ($V_{\mathrm{g}}=0.5,~6$ V). We find that the device sensitivity is better at higher gate voltages. At low $V_{\mathrm{g}}$'s the device is lossy as Andreev bound states are not properly formed due to low density and transparency. Consequently, it becomes less sensitive to heating effects.
	}}
\end{figure}

Here, we discuss the bolometer response for different gate voltages, see Fig.~\ref{fig:bolosidebndVg}. We find that the device sensitivity is better at higher gate voltages. At low $V_{\mathrm{g}}$'s the device is a bit lossy as Andreev bound states are not properly formed due to low density and transparency. Consequently, it becomes less sensitive to heating effects.

\vspace{0.3cm}

\section{Low-frequency heater modulation experiments probed using VNA}\label{sec:bolotimedomainVNA}
\begin{figure}[H]
	\centering
	\includegraphics[width=15cm]{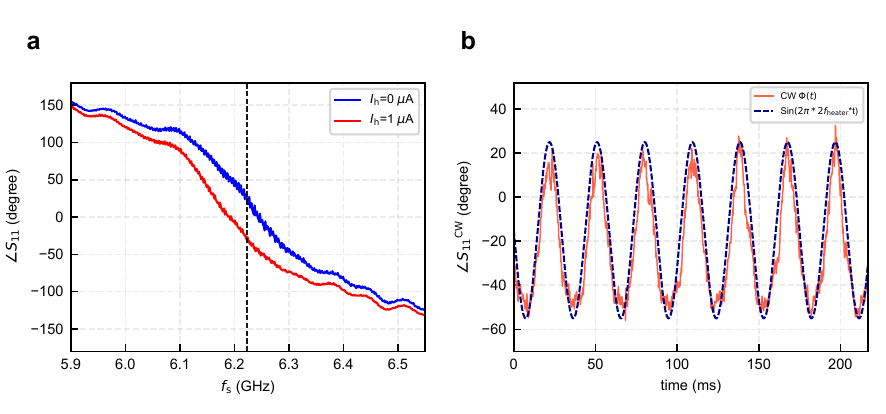}
	\caption[Low-frequency heater modulation experiments probed using VNA]{ \label{fig:boloCWoscll} {\textbf{Low-frequency heater modulation experiments probed using VNA.}}
		{
			\textbf{a,}~Shows phase response ($\angle S_{11}$) of the bolometer for a heater ON (red)/ OFF (blue) situation. 
			\textbf{b,}~Shows the oscillation in the phase of a pump tone parked at $6.22$ GHz, shown by a dashed vertical line in Fig.~\ref{fig:boloCWoscll}a. Here we do a modulation of the heater current at a frequency $f_{\mathrm{h}}=17$ Hz and measure the continuous wave phase response on the VNA at the pump tone frequency $6.22$ GHz. We keep the VNA on large IFBW $\sim 1$ kHz, primarily for fast data taking. The phase of the pump tone oscillates with time between two values. The two values correspond to the heated/unheated phase points in Fig.~\ref{fig:boloCWoscll}a. A sinusoid fit yields an oscillation frequency of $2f_{\mathrm{h}}=34$ Hz, suggesting a modulated heating of the device. 
	}}
\end{figure}

Here we discuss the heater modulation experiments probed using a VNA before employing a spectrum analyzer. Typically, VNA measurements are slow due to the filtering time involved in the phase-locked lock-in scheme. However, for observing millisecond range time domain data, a VNA can capture the signal quite accurately. For fast time-domain measurements in the sub-millisecond range, fast ADC cards are preferred over VNA. For a similar reason, we used the Alazar ATS9870 for fast-time domain measurements. We measured the oscillation in the pump tone's phase as a function of time while doing the heater modulation. See Fig.~\ref{fig:boloCWoscll},  where we present data from device-1 during our initial checks.
\vspace{0.3cm}

\section{Mixing plate heating vs on flake heating response}\label{sec:bolomixpltvsheater}
\vspace{0.3cm}
\begin{figure}[H]
	\centering
	\includegraphics[width=15cm]{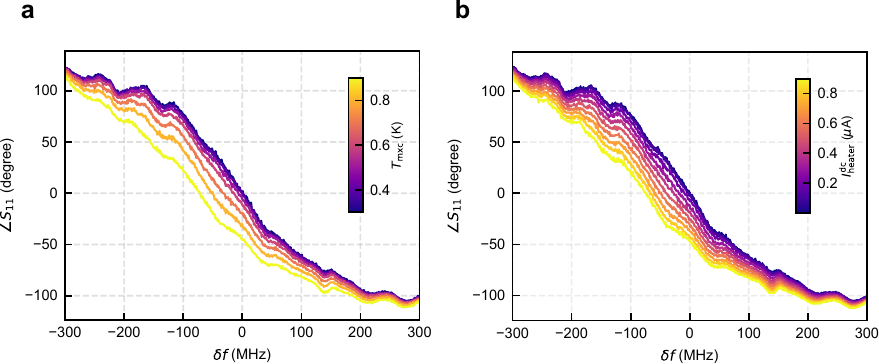}
	\caption[Mixing plate heating vs on flake heating response]{ \label{fig:boloheatvsmxplat} {\textbf{Mixing plate heating vs on flake heating response.}}
		{
			\textbf{a,}~Shows the phase response ($\angle S_{11}$) of the device as a function of signal frequency when we vary the bath temperature ($T_{\mathrm{mxc}}$) and keep the heater current at zero. This is taken at a fixed gate voltage $V_{\mathrm{g}}=6$ V, in low powers. The x-axis represents the detuning frequency from the unheated resonance frequency.
			\textbf{b,}~Shows the phase response ($\angle S_{11}$) of the device as a function of signal frequency for different heater currents ($I_{\mathrm{heater}}^{\mathrm{dc}}$) while we kept the mixing plate temperature ($T_{\mathrm{mxc}}$) constant at $20$ mK. This is taken at a fixed gate voltage $V_{\mathrm{g}}=6$ V, in low powers. The x-axis represents the detuning frequency from the unheated resonance frequency.
	}}
\end{figure}
Here we show the heating effect in our bolometer for mixing plate heating vs on flake heating scheme, see Fig.~\ref{fig:boloheatvsmxplat}. We find that the significant shift in the phase ($\angle S_{11}$) occurs after $400$ mK while increasing the mixing plate temperature, suggesting the bolometer is not very sensitive for sub $100$ mK temperatures. This can happen because of the large proximity effect of MoRe, which has a large superconducting energy gap $\sim2.3$ meV and $T_{\mathrm{c}}\sim9$ K. In the future, low $T_{\mathrm{c}}$ or gap superconductors like Al or $\beta$-Ta can be used for the superconducting contacts to enhance the sensitivity further.

\section{Finite element simulation for estimating the heat transfer to phonons}
To support our observation on phonon-based heat dissipation to bath, we perform COMSOL finite element simulations. We solved the coupled heat transfer and electric currents multiphysics model for the graphene JJ connected to a Joule heater. We use an experimental lookup table based on the supplementary Fig.~\ref{fig:boloheatvsmxplat}, where we get the graphene electron temperature as a function of the Joule heating current. This data set is used to verify the observations from finite element simulations. The system reaches a very high steady-state temperature, without considering the electron-phonon coupling in the simulation (see supplementary Fig.~\ref{fig:heating_trends2}a). The simulated temperature matches the experimental dataset trend by considering the e-p coupling term, which is computed across all the layers in the system. A significant fraction of the heat ($\sim 35\%$) generated by the heater leaks to the bath via phonons, as shown in supplementary Fig.~\ref{fig:heating_trends2}b. The PDE governing the heat propagation in the graphene flake can be written as:
\begin{equation*}
    c_{\mathrm{e}}\frac{\partial{T_e}}{\partial{t}} = \nabla\cdot(k_e \nabla T_e)- \Sigma_{\mathrm{e-p}} (T_e^\delta - T_0^\delta)
\end{equation*}
The term on the left-hand side is the rate of internal energy change of the hot electrons. The first term on the right-hand side describes the heat propagation through the electron diffusion and the second term is the heat dissipation due to the e-p coupling, where the $\Sigma_{\mathrm{e-p}}$ is the e-p coupling parameter, $\delta$ is the temperature power-law exponent, and $T_{e}$ and $T_{0}$ are the electron and bath temperatures respectively~\cite{walsh_graphene-based_2017,fried_performance_2024}. At the steady state of the device, the time-dependent term goes to zero. The device model used for the COMSOL simulations is shown in supplementary Fig.~\ref{fig:heating_trends1}. This structure is constructed using the geometrical and material parameters detailed in Table 1 $\&$ 2. The graphene layer is trenched to incorporate the superconducting JJ and the heater electrodes into the structure. Here, the JJ arms are modeled as superconductors with very high conductivity and the heat transfer equation is not computed within the arms, considering them as thermal insulators. In our scaled model, the graphene JJ has a gap of 1 $\mu$m between the superconducting electrodes. The SiO$_2$ is modeled as a near-ideal heat sink.

\vspace{0.3cm}
\begin{figure}[H]
    \centering
    \includegraphics[width=15cm]{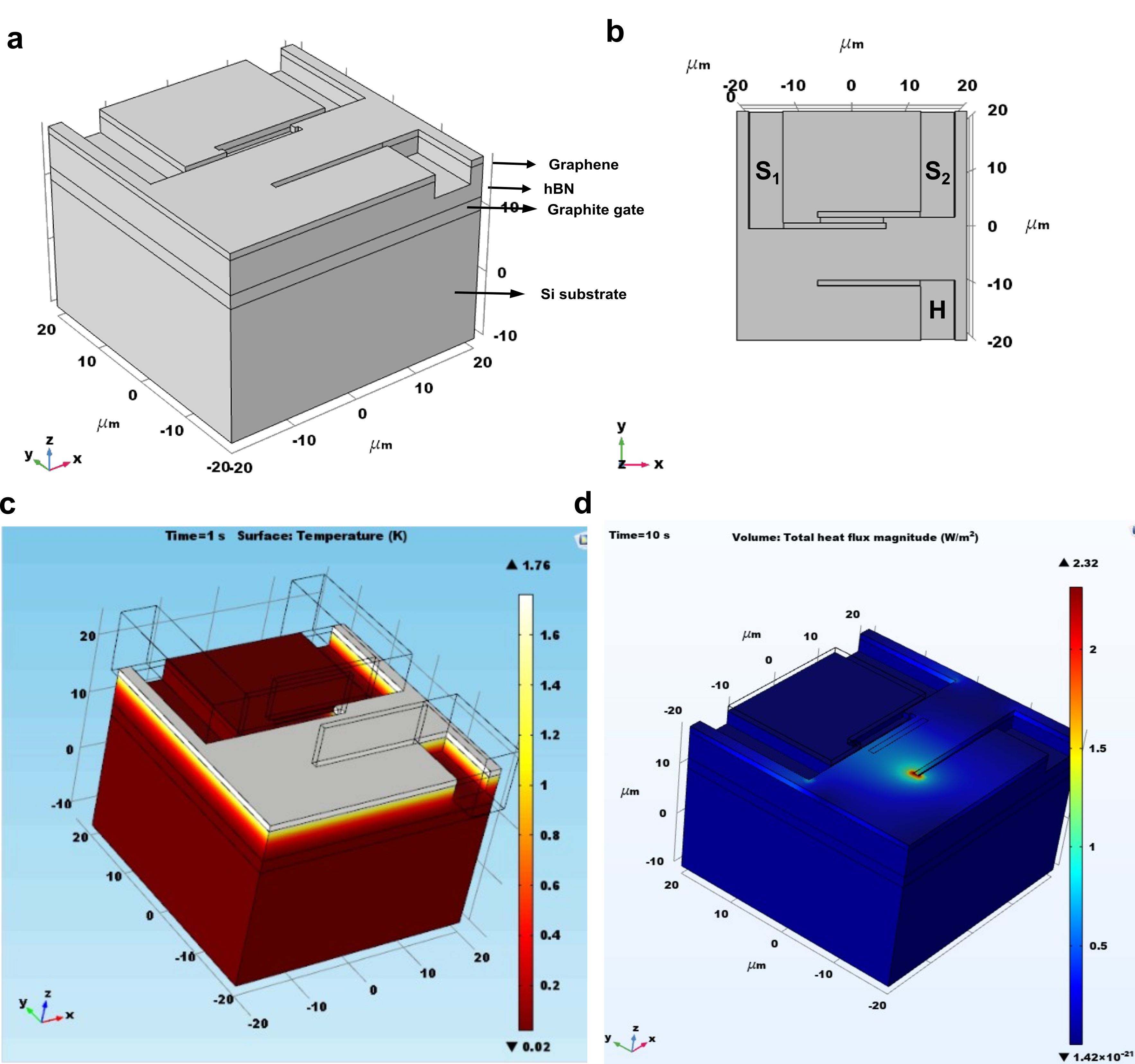}
    \caption[Device model in COMSOL simulation and heating effects]{\label{fig:heating_trends1}{\textbf{Device model in COMSOL simulation and heating effects.}}
    {\textbf{a}, Shows the side-angled view of the model structure, where the top layer is the graphene, followed by hBN, graphite gate, and Si substrate. \textbf{b}, Shows the top view of the graphene JJ with heater electrode. The electrodes are modeled as a trench through graphene, to make the JJ superconducting leads (S$_{1}$ $\&$ S$_{2}$), and heater line(H). \textbf{c}, Shows the steady-state temperature profile across the layers, for a specific heating current. \textbf{d}, Shows the heat flux propagation, generated by the Joule heating, throughout the system in the steady state.}}
    \label{fig:enter-label}
\end{figure}

The equations used in the Electric Currents and Heat Transfer in Solids module are as follows:\\
Electric Currents [ec]:
\begin{equation*}
\begin{split}
    \nabla\cdot J & = Q_{j,v}\\
    J & = \sigma E + \frac{\partial{D}}{\partial t} + J_e\\
    E & =  -\nabla V
\end{split}  
\end{equation*}
Heat transfer in Solids [ht]:
\begin{equation*}
    \rho C_p\frac{\partial{T}}{\partial{t}} = \nabla\cdot(k\nabla T) + Q_e
\end{equation*}
To incorporate the electron-phonon coupling term, the coupling was introduced in the heat transfer equation as "$-\Sigma_{\mathrm{e-p}}(T^4-T_0^4)$", by modifying the term Q$_{e}$ in the above equation in COMSOL. Here, the negative sign in front of the equation indicates that the e-p term removes heat from the system, effectively acting as a heat sink. For the purpose of this study, the arms of the Josephson junction were grounded, and direct current (DC) was supplied to the heater arm. Steady-state temperature values were recorded for each corresponding value of the heater current.

\vspace{0.3cm}
\begin{figure}[H]
    \centering
    \includegraphics[width=15cm]{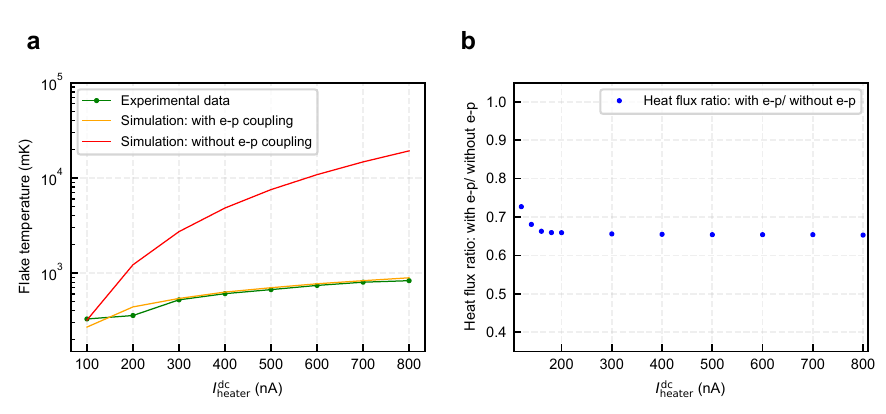}
    \caption[Electron temperature with and without e-p coupling, and heat flux ratio with and without e-p coupling]{\label{fig:heating_trends2}{\textbf{Electron temperature with and without e-p coupling, and heat flux ratio with and without e-p coupling.}}
    {\textbf{a}, Shows the electron temperature obtained from COMSOL simulation without factoring the e-p coupling significantly deviating from the experimental temperature trend. However, the simulation trend follows the experimental temperature trend on adding the e-p coupling term, which accounts for the heat dissipation to the substrate phonons. 
    \textbf{b}, Shows the ratio of heat flux near the JJ, with and without factoring in the e-p coupling, which is $\sim0.65$.}}
    \label{fig:enter-label}
\end{figure}
\begin{table}[h]
    \centering
    \begin{tabular}{|l|c|c|c|c|c|}
        \hline
        \thead{Parameters} & \thead{Graphene \\(Literature)} & \thead{Graphene \\(Simulation)} & \thead{hBN \\(Literature)} & \thead{hBN \\(Simulation)} \\
        \hline
        Heat capacity & $5.1 \times 10^{-6}$ J/kg K & 8 J/kg K & 770 J/kg K & 20 J/kg K \\
        \hline
        Thermal conductivity & $3 \times 10^{-4}$ W/mK & 100 W/m K & 19 W/m K & $10^{-10}$ W/m K \\
        \hline
        e-p coupling const $\Sigma$ (Lit.) & 0.2 W/K$^4$ m$^2$ & & & \\
        \hline
        e-p coupling const $\Sigma$ (Sim.) & $0.6 \times 10^{6}$ W/K$^4$ m$^2$ & & & \\
        \hline
    \end{tabular}
    \caption{Thermal properties of graphene $\&$ hBN}
    \label{tab:graphene_hbn_parameters}
\end{table}

\begin{table}[h]
    \centering
    \begin{tabular}{|c|c|}
    \hline
    \textbf{Materials} & \textbf{Dimensions (Simulation)} \\ \hline
    Graphene & 40 um x 40 um x 1 um \\ \hline
    Bottom hBN & 40 um x 40 um x 6 um \\ \hline
    Graphite gate & 40 um x 40 um x 2 um \\ \hline
    Si substrate & 40 um x 40 um x 20 um \\ \hline
    \end{tabular}
    \caption{Material dimensions in the COMSOL simulation setup.}
    \label{tab:material_dimensions}
\end{table}

\end{document}